\documentclass[runningheads]{llncs}
\usepackage{graphicx}

\usepackage{wrapfig}
\usepackage{amsmath,amssymb}
\usepackage{stmaryrd,mathrsfs}

\makeatletter
\RequirePackage[bookmarks,unicode,colorlinks=true]{hyperref}\def\@citecolor{blue}\def\@urlcolor{blue}\def\@linkcolor{blue}
\def\orcidID#1{\smash{\href{http://orcid.org/#1}{\protect\raisebox{-1.25pt}{\protect\includegraphics{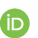}}}}}
\makeatother

\usepackage[strings]{underscore}

\usepackage{xcolor}

\usepackage[appendix=append]{apxproof}

\spnewtheorem{mydefinition}[definition]{Definition}{\bfseries}{\rmfamily}
\newtheoremrep{mytheorem}[definition]{Theorem}
\newtheoremrep{myproposition}[definition]{Proposition}
\newtheoremrep{mycorollary}[definition]{Corollary}
\newtheoremrep{myconjecture}[definition]{Conjecture}
\spnewtheorem{myexample}[definition]{Example}{\itshape}{\rmfamily}
\spnewtheorem{myassumption}[definition]{Assumption}{\bfseries}{\rmfamily}
\spnewtheorem{myquestion}[definition]{Question}{\bfseries}{\rmfamily}
\spnewtheorem{myfact}[definition]{Fact}{\bfseries}{\rmfamily}
\spnewtheorem{myremark}[definition]{Remark}{\itshape}{\rmfamily}
\spnewtheorem*{myclaim}{Claim}{\bfseries}{\rmfamily}

\usepackage{comment}
\includecomment{auxproof}

\usepackage[all]{xy}
\newdir{ >}{{}*!/-8pt/@{>}}

\newcommand{\CC}{\mathbb C}
\newcommand{\DD}{\mathbb D}
\newcommand{\EE}{\mathbb E}
\newcommand{\FF}{\mathbb F}

\newcommand{\RR}{\mathbb R}

\newcommand{\Id}{\mathrm{Id}}
\newcommand{\id}{\mathrm{id}}
\newcommand{\co}{\mathbin{\circ}}
\newcommand{\op}{\mathrm{op}}

\newcommand{\bOmega}{\mathbf{\Omega}}

\newcommand{\mclass}{\mathcal{M}}

\newcommand{\OpenSet}{\mathcal{O}}

\newcommand{\Top}{\mathbf{Top}}
\newcommand{\Set}{\mathbf{Set}}
\newcommand{\PMetT}{\mathbf{PMet}_{\top}}
\newcommand{\Pre}{\mathbf{Pre}}
\newcommand{\EqRel}{\mathbf{EqRel}}
\newcommand{\ERel}{\mathbf{ERel}}
\newcommand{\CLatw}{\mathbf{CLat}_{\sqcap}}

\newcommand{\lefib}{\sqsubseteq}
\newcommand{\gefib}{\sqsupseteq}
\newcommand{\bigwedgefib}{\bigsqcap}

\newcommand{\Pos}{\mathbf{Pos}}

\newcommand{\codlif}[3]{ {#1}^{ {#2},{#3} } }

\newcommand{\pow}{\mathcal{P}}
\newcommand{\Distle}{\mathcal{D}_{\le 1}}
\newcommand{\Sierp}{\mathbb{O}}
\newcommand{\Thr}{\mathrm{thr}}
\newcommand{\Acc}{\mathrm{acc}}
\newcommand{\Next}[1]{ \left\langle {#1} \right\rangle}

\newcommand{\codPT}[3]{ \Phi^{ {#1},{#2} }_{#3} }

\begin{document}
\title{Injective Objects and Fibered Codensity Liftings\thanks{The
    author was supported by ERATO HASUO Metamathematics for Systems
    Design Project (No. JPMJER1603), JST.
    This is the accepted manuscript of~\cite{Komorida-CMCS2020}.
    The final publication is available at Springer via \url{http://dx.doi.org/10.1007/978-3-030-57201-3_7}.}}
\author{Yuichi Komorida\inst{1,2}\orcidID{0000-0002-3371-5243}}
\authorrunning{Y. Komorida}
\institute{The Graduate University for Advanced Studies, SOKENDAI,
  Tokyo, Japan\and National Institute of Informatics, Tokyo, Japan,
  \email{komorin@nii.ac.jp}}
\maketitle 

\begin{abstract}
  Functor lifting along a fibration is used for several different purposes in
  computer science. In the theory of coalgebras, it is used to define
  coinductive predicates, such as simulation preorder and bisimilarity.  Codensity lifting is a scheme to obtain a
  functor lifting along a fibration.  It generalizes a few
  previous lifting schemes including the Kantorovich lifting.
  In this paper, we seek a property of functor lifting called fiberedness.
  Hinted
  by a known result for Kantorovich lifting, we identify a sufficient
  condition for a codensity lifting to be fibered.  We see that this
  condition applies to many examples that have been studied.  As an
  application, we derive some results on bisimilarity-like notions.
\end{abstract}

\pushQED{\qed}

\section{Introduction}
\label{sec:introduction}

In this paper, we focus on a category-theoretical gadget, called
\emph{functor lifting}, and seek a property thereof, called
\emph{fiberedness}.  As is often the case with such mathematical
topics, functor lifting comes up in several different places in
computer science under various disguises (as mentioned in
Section~\ref{subsec:relatedwork}).  Here we see one of such places,
\emph{bisimilarity and its generalizations on coalgebras}, before we
formally introduce functor lifting.

\subsection{Coalgebras and Bisimilarity}
Computer programs work as we write them, not necessarily as we expect.
One approach to overcome this gap is to \emph{verify} the systems so
that we can make sure that they meet our requirements.  Abstract
mathematical methods are often useful for the purpose, but before that,
we have to \emph{model} the target system by some mathematical
structure.

\emph{Coalgebra}~\cite{Rutten-TCS2000-coalgebra} is one of such
mathematical structure with a broad scope of application. It is
defined in terms of the theory of \emph{categories and
  functors}. Given a category $\CC$ and an endofunctor
$F\colon\CC\to\CC$, an $F$-coalgebra is defined as an arrow
$c\colon X\to FX$. This simple definition includes many kinds of
state-transition systems as special cases, e.g., Kripke frame (and
model), Markov chain (and process), and (deterministic and
non-deterministic) automata.

Having modeled a system as a coalgebra, we can ask a fundamental
question: which states behave the same?  \emph{Bisimilarity}~\cite{Milner-1989-bisimulation,Park-1981-bisimulation} is one of
the notions to define such equivalence. (For an introduction, see, e.g.,~\cite{Sangiorgi-book2011-bisimulation}.)  We sketch the idea in the case
where $\CC=\Set$ and $F=\Sigma\times(-)$.  In this case,
$F$-coalgebras are deterministic LTSs.  Consider a coalgebra
$c\colon X\to\Sigma\times X$ and define $l\colon X\to\Sigma$ and
$n\colon X\to X$ by $(l(x),n(x))=c(x)$.  The point here is the
following observation: if $x,y\in X$ behave the same, then $l(x)=l(y)$
must hold, and $n(x)$ and $n(y)$ must behave the same.  This is almost
the definition of bisimilarity: the bisimilarity relation is the
greatest binary relation $\sim\subseteq X\times X$ that satisfies
\[ x\sim y \implies l(x)=l(y)\wedge n(x)\sim n(y).
\]

For other functors $F$, the idea is roughly the same: in a coalgebra
$c\colon X\to FX$, for $x,y\in X$ to behave the same, $c(x)$ and
$c(y)$ must behave the same.  To define bisimilarity precisely,
however, we have to turn a relation $R\subseteq X\times X$ into
$R'\subseteq FX \times FX$.

\subsection{Qualitative and Quantitative Bisimilarity from Functor Lifting}
\label{sec:bisimilarityfromlifting}
An elegant way to formulate this is the following: bundle binary
relations on all sets into one \emph{fibration} and use \emph{functor
  lifting} as in~\cite{HermidaJacobs-InfComp1998}. We give ideas on
them here. The precise definitions are in
Section~\ref{sec:preliminaries}.

First, we gather all pairs $(X,R)$ of a set $X$ and a binary relation
$R\subseteq X\times X$ into one category $\ERel$
(\autoref{exam:erelpreeqrel}).  It comes with a forgetful functor
$U\colon\ERel\to\Set$.  (This is a \emph{fibration}.)  Any binary
relation $R$ on $X$ is sent to $X$ by $U$; placing the things
vertically, $R$ is ``above'' $X$.  Now let us assume that there exists a functor
$\dot{F}\colon\ERel\to\ERel$ satisfying $U\co\dot{F}=F\co U$.  This
means that any binary relation $R$ on $X$ is sent to one on $FX$:
\[
  \xymatrix{\ERel\ar[r]_{\dot{F}}\ar[d]_{U} & \ERel\ar[d]_{U} & R\ar@{..}[d]\ar@{|->}[r] & \dot{F}R\ar@{..}[d]\\
  \Set\ar[r]_{F} & \Set & X\ar@{|->}[r] & FX }
\]
(This means that the functor $\dot{F}$ is a \emph{lifting} of $F$
along $U$.)  The functor $U\colon\ERel\to\Set$ has an important
structure: for any $f\colon Y\to X$ and a relation $R$ on $X$, we can
obtain a relation $f^*R$ on $Y$ in a canonical
way:\[ f^*R = \{(y,y')\in Y\times Y | (f(y),f(y')) \in R\}.
\](This is called \emph{reindexing} or \emph{pullback}.)  By using these, we can define the
bisimulation relation on $c\colon X\to FX$ as the greatest fixed point
of $f^* \co \dot{F}$.

An advantage of this approach is that we can readily generalize this
to other ``bisimilarity-like'' notions.  For example, by changing the
fibration to $\PMetT\to\Set$ (\autoref{exam:pmettclatwfib}), one can
define a \emph{behavioral (pseudo)metric}~\cite{BaldanBKK-FSTTCS2014}.

\subsection{Codensity Lifting of Endofunctors}
Now we know that a functor lifting induces a bisimilarity-like notion.
Then, how can we obtain a functor lifting?  \emph{Codensity Lifting}
is a scheme to obtain such liftings.  It is first introduced by
Katsumata and Sato~\cite{KatsumataSato-CALCO2015} for monads
using \emph{codensity monad} construction~\cite{Leinster-TAC2013}.
It is later
extended to general endofunctors by Sprunger et
al.~\cite{SprungerKDH-CMCS2018}.  It is parametrized in a set of data
called a \emph{lifting parameter}.  By changing lifting parameters, a
broad class of functor liftings can be represented as codensity
liftings, as is shown, e.g.,~in Komorida et
al.~\cite{KomoridaKHKH-LICS2019}.

As mentioned in the last section, we can define a bisimilarity-like
notion using codensity lifting. It is called \emph{codensity
  bisimilarity} in~\cite[Sections III and VI]{KomoridaKHKH-LICS2019}.

\subsection{Fiberedness of Lifting}
In some situations, we have to assume that $\dot{F}\colon\EE\to\EE$
interacts well with the pullback operation between the fibers. In such a
situation, $\dot{F}$ is required to be \emph{fibered} (\autoref{def:fiberedness}).  It
means that pullbacks and $\dot{F}$ are ``commutative,'' in the sense
that they satisfy $\dot{F}(f^*P)=(Ff)^*(\dot{F}P)$.

Some of the existing works indeed require fiberedness. For example,
Hasuo et al.~\cite[Definition 2.2]{HasuoCKJ-ENTCS2013} include
fiberedness in their definition of predicate lifting.  Fiberedness
also plays a notable role in~\cite{BaldanBKK-LMCS2018}, where it is
rephrased to isometry-preservation.  However, there has been no
systematic result on fiberedness of codensity lifting.

\subsection{Contributions}
In the current paper, hinted by a result of Baldan et
al.~\cite{BaldanBKK-LMCS2018}, we show a sufficient condition on the
lifting parameter guaranteeing the resulting functor to be fibered
(\autoref{thm:fiberednessfrominjective}).  The scope of our
fiberedness result is so broad that it covers, e.g., most of the
examples presented in~\cite{KomoridaKHKH-LICS2019}
(Section~\ref{sec:examples}).

The condition involves a variation of the notion of injective object,
which we call \emph{c-injective object}
(\autoref{def:cinjectiveobject}).  To our knowledge, such a notion
connecting injective objects and fibrations is new.  We study some
basic properties of them.

Using the fiberedness result, we show a property of codensity
bisimilarity which we call \emph{stability under coalgebra morphisms} (\autoref{prop:coalgebramorphismiscartesian}).
As a corollary, we see that, when there is a final coalgebra, the codensity
bisimilarity on any coalgebra is determined by that on the final
coalgebra.  Note that this kind of property is well-known for a
conventional bisimilarity relation (\autoref{cor:codenbisimfromfinalcoalg}).

To summarize, our technical contributions are as follows:
\begin{itemize}
\item We define \emph{c-injective objects} for fibrations
  (\autoref{def:cinjectiveobject}) and show some properties of them.
\item We show a sufficient condition on the lifting parameter to
  guarantee fiberedness of codensity lifting
  (\autoref{thm:fiberednessfrominjective} and
  \autoref{cor:fiberednessfrominjectivemulti}).
\item We show a number of examples (Section~\ref{sec:examples}) to which
  the condition above is applicable.
\item As an application, we show that codensity bisimilarity is
  stable under coalgebra morphisms (\autoref{prop:coalgebramorphismiscartesian}) in many cases, including a
  new one (\autoref{exam:bisimtopfromfinalcoalg}).
\end{itemize}

\subsection{Related Work}
\label{subsec:relatedwork}
Even though we focused on bisimilarity and coalgebra above, functor lifting comes up in
computer science here and there.  To name a few, it has applications
in logical
predicates~\cite{HermidaJacobs-InfComp1998,Katsumata-CSL2005},
quantitative bisimulation~\cite{BaldanBKK-LMCS2018}, and differential
privacy~\cite{SatoBGHK-LICS2019}.

As mentioned above, there have been many methods to obtain liftings of
functors.  \emph{Kantorovich
  lifting}~\cite{BaldanBKK-FSTTCS2014,KoenigMikaMichalski-CONCUR2018}
and \emph{generalized Kantorovich
  metric}~\cite{ChatzikokolakisGPX-CONCUR2014} are both special cases
of the version of codensity lifting considered here.
\emph{Categorical $\top\top$-lifting}~\cite{Katsumata-CSL2005} is
the precursor of the original version of
codensity lifting, but it is not a special case of codensity lifting.
For categorical $\top\top$-lifting, one uses internal Hom-objects rather than Hom-sets like codensity lifting. Obtaining a sufficient condition for fiberedness of categorical $\top\top$-lifting is future work.
\emph{Wasserstein lifting}~\cite{BaldanBKK-FSTTCS2014} is another
method that is somehow dual to Kantorovich lifting.  They have shown
that any lifting obtained by this scheme is fibered.
Klin~\cite{Klin-CALCO2005} goes a different way: rather than showing
fiberedness, they incorporate fiberedness in the definition.  They
study \emph{the least fibered lifting} along $\EqRel\to\Set$ and show
that, in good situations, it coincides with the \emph{canonical
  relation lifting}.

The notion of \emph{injective object} is first introduced in
homological algebra as \emph{injective
  modules}~\cite{Baer-1940-injectivemodule}.  There are also some
works about injective objects outside homological algebra:
Scott~\cite{Scott-1972-contlattice} and Banaschewski and
Bruns~\cite{BanaschewskiBruns-1967} have identified the injective
objects in $\Top_0$ and $\Pos$, respectively. We use their results in
Section~\ref{sec:moreoncinjectives} (where the categories mentioned are defined).  Injective objects
w.r.t.~isometric embeddings in the category of metric spaces are also
well-studied and called \emph{hyperconvex
  spaces}~\cite{EspinolaKhamsi-2001-hyperconvex}. Finding a precise
connection between them and c-injective objects in $\PMetT\to\Set$ (\autoref{exam:pmettclatwfib}) is
future work. Recently, in his preprint~\cite{Fujii-arXiv2019}, Fujii
has extended the above result in $\Pos$ and characterized injective
objects in the category of $\mathcal{Q}$-categories with respect to
the class of fully faithful $\mathcal{Q}$-functors, for any quantale
$\mathcal{Q}$.

\subsection{Organization}
In Section~\ref{sec:preliminaries}, we review
\emph{$\CLatw$-fibrations} and \emph{functor liftings}.  In Section~
\ref{sec:mainthings}, we review the definition of \emph{codensity
  lifting} and introduce the notion of \emph{c-injective objects}. We
show a sufficient condition for a codensity lifting to be fibered.  In
Section~\ref{sec:moreoncinjectives}, we show some general results on
c-injective objects.  In Section~\ref{sec:examples}, we list several
examples of fibered codensity liftings using the results in
Section~\ref{sec:moreoncinjectives}.  In
Section~\ref{sec:applications}, we apply the fiberedness result to
\emph{codensity bisimilarity}.  In Section~\ref{sec:conclusions}, we
conclude with some remarks and future work.

\section{Preliminaries}
\label{sec:preliminaries}

We assume some knowledge of \emph{category theory}, but the full content of
the standard reference~\cite{Maclane-book1978-CWM} is not needed.  The
basic definitions and theorems, e.g., those in
Leinster~\cite{Leinster-book2014}, is enough.  Even though we have
explained our motivation through coalgebra, no knowledge of coalgebra
is needed for the main result in Section~\ref{sec:mainthings}.

In the following, $\Set$ means the category of sets and
(set-theoretic) functions.

\subsection{$\CLatw$-Fibrations}

Here we introduce \emph{$\CLatw$-fibrations}, as defined
in~\cite{KomoridaKHKH-LICS2019}. We use them to model various
``notions of indistinguishability'' like preorder, equivalence
relation, and pseudometric. Assuming full knowledge of the theory of
fibrations, we could define them as poset fibrations with fibered
small meets.  Instead, we give an explicit definition below.  This is
mainly because we need the notion of \emph{Cartesian arrow}.  For a
comprehensive account of the theory of fibrations, the reader can consult, e.g., a book
by Jacobs~\cite{Jacobs-book1999-CLTT} or Hermida's
thesis~\cite{Hermida-thesis1993}, but
in the following, we do not assume any knowledge of fibrations.

We first define a fiber of a functor over an object.
Basically, this is only considered in the case where the functor is a fibration.

\begin{mydefinition}[fiber]
  Let $p\colon\EE\to\CC$ be a functor and $X\in\CC$ be an object.
  The \emph{fiber over} $X$ is the subcategory of $\EE$
  \begin{itemize}
  \item whose objects are $P\in\EE$ such that $pP=X$ and
  \item whose arrows are $f\colon P\to Q$ such that $pf=\id_X$.
  \end{itemize}
  We denote it by $\EE_X$.
\end{mydefinition}

Note that, if $p$ is faithful, then each fiber is a thin category, i.e., a preorder.

The following definition of poset fibration is a special case of that in~\cite{Jacobs-book1999-CLTT}.

\begin{mydefinition}[cartesian arrow and poset fibration]
  \label{def:cartesian-posetfibration}
  Let $p\colon\EE\to\CC$ be a faithful functor.
  
  An arrow $f\colon P\to Q$ in $\EE$ is \emph{Cartesian} if the following condition is satisfied:
  \begin{itemize}
  \item For each $R\in\EE$ and $g\colon R\to Q$, there exists
    $h\colon R\to P$ such that $g=f\co h$ if and only if there exists
    $h'\colon pR\to pP$ such that $pg=pf \co h'$.
  \end{itemize}

  The functor $p$ is called a \emph{poset fibration} if the following are satisfied:
  \begin{itemize}
  \item For each $X\in\CC$, the fiber $\EE_X$ is a poset. The order is
    denoted by $\lefib$. We define the direction so that $P\lefib Q$
    holds if and only if there is an arrow $P\to Q$ in $\EE_X$.
  \item For each $Q\in\EE$ and $f\colon X\to pQ$, there exists an
    object $f^*Q\in\EE_X$ and a Cartesian arrow
    $\dot{f}\colon f^*Q\to Q$ such that $p\dot{f}=f$. (Such $f^*Q$ and
    $\dot{f}$ are necessarily unique.)
  \end{itemize}
  The map $Q\mapsto f^*Q$ turns out to be a monotone map from $\EE_Y$ to $\EE_X$.
  We call it the \emph{pullback functor} and denote it by $f^*\colon\EE_Y\to\EE_X$.
\end{mydefinition}

Intuitively, pullback functors model substitutions.  Indeed, in many
examples, they are just ``assigning $f(x)$ to $y$'', as can be seen
below.

\begin{myexample}[pseudometric]
  \label{exam:pmettfib}
  Let $\top$ be a positive real number or $+\infty$.
  Define a category $\PMetT$ as follows:
  \begin{itemize}
  \item Each object is a pair $(X,d)$ of a set $X$ and a
    $[0,\top]$-valued pseudometric $d\colon X\times X\to [0,\top]$. (A
    pseudometric is a metric without the condition
    $d(x,y)=0\implies x=y$.)
  \item Each arrow from $(X,d_X)$ to $(Y,d_Y)$ is a nonexpansive map
    $f\colon X\to Y$. ($f$ is nonexpansive if, for all $x$ and
    $x'\in X$, $d_X(x,x')\ge d_Y(f(x),f(x'))$.)
  \end{itemize}
  The obvious forgetful functor $\PMetT\to\Set$ is a poset
  fibration.  For each $X\in\Set$, the objects of the fiber
  $(\PMetT)_X$ are the pseudometrics on $X$.  However, the order is
  reversed: in our notation, the order is defined by
  \[ (X,d_1)\lefib(X,d_2) \Leftrightarrow \forall x,x'\in X,
    d_1(x,x')\ge d_2(x,x').
  \]

  An arrow $f\colon (X,d_X)\to(Y,d_Y)$ is Cartesian if and only if it
  is an isometry, i.e., $d_X(x,x')= d_Y(f(x),f(x'))$ holds for all
  $x,x'$.  For $(Y,d_Y)\in\PMetT$ and $f\colon X\to Y$, the pullback
  $f^*(Y,d_Y)$ is the set $X$ with the pseudometric
  $(x,x')\mapsto d_Y(f(x),f(x'))$.
\end{myexample}

We list a few properties of pullback functors that we use:
\begin{myproposition}
  \label{prop:decencyandpullback}
  Let $p\colon\EE\to\CC$ be a poset fibration, $f\colon X\to Y$ be an
  arrow in $\CC$ and $P\in\EE_X$ and $Q\in\EE_Y$ be objects in $\EE$.
  There exists an arrow $g\colon P\to Q$ such that $pg=f$ if and only
  if $P\lefib f^*Q$.  Moreover, such $g$ is Cartesian if and only if
  $P=f^*Q$.\qed
\end{myproposition}

\begin{myproposition}
  \label{prop:pullbackisfunctorial}
  Let $p\colon\EE\to\CC$ be a poset fibration.
  \begin{itemize}
  \item For each $X\in\CC$, $(\id_X)^*\colon\EE_X\to\EE_X$ is the
    identity functor.
  \item For each composable pair of arrows
    $X\xrightarrow{f}Y\xrightarrow{g}Z$ in $\CC$,
    $(g\co f)^* = f^* \co g^*$ holds.\qed
  \end{itemize}
\end{myproposition}

Now we define the class that we are concerned about, $\CLatw$-fibrations.

\begin{mydefinition}[$\CLatw$-fibration]
  A poset fibration $p\colon\EE\to\CC$ is a \emph{$\CLatw$-fibration} if the following conditions are satisfied:
  \begin{itemize}
  \item Each fiber $\EE_X$ is small and has small meets, which we denote by $\bigwedgefib$.
  \item Each pullback functor $f^*$ preserves small meets.
  \end{itemize}
\end{mydefinition}

Note that, in the situation above, each fiber $\EE_X$ is a complete
lattice: the small joins can be constructed using small meets.

\begin{myexample}[pseudometric]
  \label{exam:pmettclatwfib}
  The poset fibration $\PMetT\to\Set$ in \autoref{exam:pmettfib} is a
  $\CLatw$-fibration.  Indeed, meets can be defined by sups of
  pseudometrics: if we let $(X,d)=\bigwedgefib_{a\in A}(X,d_a)$, then
  \[ d(x,x') = \sup_{a\in A} d_a(x,x')
  \] holds.
\end{myexample}

\begin{myexample}[binary relations]
  \label{exam:erelpreeqrel}
  Define a category $\ERel$ of sets with an endorelation as follows:
  \begin{itemize}
  \item Each object is a pair $(X,R)$ of a set $X$ and a binary
    relation $R\subseteq X\times X$.
  \item Each arrow from $(X,R_X)$ to $(Y,R_Y)$ is a map
    $f\colon X\to Y$ preserving the relations; that is, we require $f$
    to satisfy $(x,x')\in R_X \implies (f(x),f(x'))\in R_Y$.
  \end{itemize}
  The obvious forgetful functor $\ERel\to\Set$ is a $\CLatw$-fibration.
  For each $X\in\Set$, the fiber $\ERel_X$ is the complete lattice of
  subsets of $X\times X$.

  An arrow $f\colon (X,R_X)\to(Y,R_Y)$ is Cartesian if and only if it
  reflects the relations, i.e.,
  $(x,x')\in R_X\Leftrightarrow (f(x),f(x'))\in R_Y$ holds for all
  $x,x'$.  For $(Y,R_Y)\in\ERel$ and $f\colon X\to Y$, the pullback
  $f^*(Y,R_Y)$ is the set $X$ with the relation
  $\{(x,x')\in X\times X | (f(x),f(x')) \in R_Y\}$.

  Define the following full subcategories of $\ERel$:
  \begin{itemize}
  \item The category $\Pre$ of preordered sets and monotone maps.
  \item The category $\EqRel$ of sets with an equivalence relation and maps preserving them.
  \end{itemize}
  The forgetful functors $\Pre\to\Set$ and $\EqRel\to\Set$ are also $\CLatw$-fibrations.
\end{myexample}

$\CLatw$-fibrations are not necessarily ``relation-like''.  There also
is an example with a much more ``space-like'' flavor.

\begin{myexample}
  The forgetful functor $\Top\to\Set$ from the category $\Top$ of
  topological spaces and continuous maps is a $\CLatw$-fibration.
\end{myexample}

\subsection{Lifting and Fiberedness}

Another pivotal notion in the current paper is \emph{functor lifting}.
In Section~\ref{sec:bisimilarityfromlifting} we have seen that it is
used to define bisimilarity, or more generally bisimilarity-like
notions, as a way to turn a relation (or pseudometric, etc.) on $X$
into one on $FX$. Here we review the formal definition in a restricted
form that only considers $\CLatw$-fibration. (Note that, usually it is
defined more generally, and there are indeed applications of such
general definition.)

\begin{mydefinition}[lifting of endofunctor]
  \label{def:lifting}
  Let $p\colon\EE\to\CC$ be a $\CLatw$-fibration and
  $F\colon\CC\to\CC$ be a functor.  A \emph{lifting} of $F$ along $p$
  is a functor $\dot{F}\colon\EE\to\EE$ such that $p\co\dot{F}=F\co p$
  holds:\[
    \xymatrix{\EE\ar[r]_{\dot{F}}\ar[d]_{p} & \EE\ar[d]_{p}\\
      \CC\ar[r]_{F} & \CC. }
  \]
\end{mydefinition}

We then define \emph{fiberedness} of a lifting. This means that the
lifting interacts well with the pullback structure of the fibration,
but we first give a definition focusing on Cartesian arrows. Here we
define it in a slightly more general way so that we can use them
later (Section~\ref{sec:moreoncinjectives}).

\begin{mydefinition}[{fibered functor~\cite[Definition
  1.7.1]{Jacobs-book1999-CLTT}}]
  \label{def:fiberedness}
  Let $p\colon\EE\to\CC$ and $q\colon\FF\to\DD$ be
  $\CLatw$-fibrations.  A \emph{fibered functor} from $p$ to $q$ is a
  functor $\dot{F}\colon\EE\to\FF$ such that there is another functor
  $F\colon\CC\to\DD$ satisfying $q\co\dot{F}=F\co p$ and $\dot{F}$
  sends each Cartesian arrow to a Cartesian arrow.
\end{mydefinition}

Note that, in the situation above, such $F$ is uniquely determined by $p$, $q$, and $\dot{F}$.

Now we see a characterization of fiberedness by means of pullback.
\begin{myproposition}
  \label{prop:fiberednesspullback}
  Let $p\colon\EE\to\CC$ and $q\colon\FF\to\DD$ be $\CLatw$-fibrations
  and $\dot{F}\colon\EE\to\FF$ and $F\colon\CC\to\DD$ be functors
  satisfying $q\co\dot{F}=F\co p$.  $\dot{F}$ is a fibered functor if
  and only if, for any $f\colon X\to Y$ in $\CC$ and $P\in\EE_Y$,
  $\dot{F}(f^*P)=(Ff)^*(\dot{F}P)$ holds.\qed
\end{myproposition}
We use this in the proof of the main result.

\section{C-injective Objects and Codensity Lifting}
\label{sec:mainthings}

\subsection{Codensity Lifting}

Before we formulate our main result, we introduce \emph{codensity
  lifting} of
endofunctors~\cite{KatsumataSato-CALCO2015,SprungerKDH-CMCS2018}.
Here we use an explicit definition for a narrower situation than the
original one.

\begin{mydefinition}[codensity lifting (as
  in~\cite{KomoridaKHKH-LICS2019})]
  \label{def:codensitylifting}
  Let
  \begin{itemize}
  \item $p\colon\EE\to\CC$ be a $\CLatw$-fibration,
  \item $F\colon\CC\to\CC$ be a functor,
  \item $\bOmega\in\EE$ be an object above $\Omega\in\CC$, and
  \item $\tau\colon F\Omega\to\Omega$ be an $F$-algebra.
  \end{itemize}

  Define a functor $\codlif{F}{\bOmega}{\tau}\colon\EE\to\EE$, which is a lifting of $F$ along $p$, by
  \[ \codlif{F}{\bOmega}{\tau}P =
    \bigwedgefib_{f\in\EE(P,\bOmega)}(F(pf))^* \tau^* \bOmega
  \] for each $P\in\EE$.  The functor $\codlif{F}{\bOmega}{\tau}$ is
  called a \emph{codensity lifting} of $F$.
  Note that, for each $P\in\EE$ and $f\colon P\to\bOmega$, the situation is as follows: \[
    \xymatrix{
      & & \bOmega\ar@{..}[d] \\
      FpP\ar[r]_{F(pf)} & F\Omega\ar[r]_{\tau} & \Omega
    }
  \]
  and we can indeed obtain the pullback $(F(pf))^* \tau^* \bOmega$.
\end{mydefinition}

We have given only the object part of $\codlif{F}{\bOmega}{\tau}$
above, but the arrow part, if it is well-defined, should be determined
uniquely since $p$ is faithful.  We give a proof that it is indeed
well-defined. For each $f\colon P\to Q$, we need another arrow
$g\colon \codlif{F}{\bOmega}{\tau}P\to\codlif{F}{\bOmega}{\tau}Q$ such
that $pg=F(pf)$. By \autoref{prop:decencyandpullback}, it suffices to
show the following proposition:
\begin{myproposition}
  \label{prop:codensityliftingisfunctor}
  For any $f\colon P\to Q$,
  $\codlif{F}{\bOmega}{\tau}P \lefib (F(pf))^* \left(
    \codlif{F}{\bOmega}{\tau}Q \right)$ holds.
\end{myproposition}
\begin{proof}
  By definition, the l.h.s. satisfies
  \[ \codlif{F}{\bOmega}{\tau}P =
    \bigwedgefib_{g\in\EE(P,\bOmega)}(F(pg))^* \tau^* \bOmega.
  \]
  On the other hand, the r.h.s. satisfies
  \begin{align*}
    (F(pf))^* \left( \codlif{F}{\bOmega}{\tau}Q \right) &= (F(pf))^* \left( \bigwedgefib_{h\in\EE(Q,\bOmega)}(F(ph))^* \tau^* \bOmega \right) \\
                                                        &= \bigwedgefib_{h\in\EE(Q,\bOmega)} (F(pf))^* (F(ph))^* \tau^* \bOmega \\
                                                        &= \bigwedgefib_{h\in\EE(Q,\bOmega)} (F(p(h \co f)))^* \tau^* \bOmega.
  \end{align*}
  Here, since
  $ \left\{ g\in\EE(P,\bOmega) \right\} \supseteq \left\{ h\co f ~|~
      h\in\EE(Q,\bOmega)\right\}
  $ holds, we have
  \[ \bigwedgefib_{g\in\EE(P,\bOmega)}(F(pg))^* \tau^* \bOmega \lefib
    \bigwedgefib_{h\in\EE(Q,\bOmega)} (F(p(h \co f)))^* \tau^*
    \bOmega.
  \]
  This means
  $\codlif{F}{\bOmega}{\tau}P \lefib (F(pf))^* \left(
    \codlif{F}{\bOmega}{\tau}Q \right)$.\qed
\end{proof}

\subsection{C-injective Object}

In the proof of the functoriality of $\codlif{F}{\bOmega}{\tau}$, ultimately we use the fact that, for any
$f\colon P\to Q$, any ``test'' $k\colon Q\to \bOmega$ can be turned
into another ``test'' $k\co f\colon P\to\bOmega$.  On the other hand,
when we try to prove fiberedness of $\codlif{F}{\bOmega}{\tau}$, we
have to somehow lift a ``test'' $g\colon P\to\bOmega$ along a
Cartesian arrow $f\colon P\to Q$ and obtain another ``test''
$h\colon Q\to\bOmega$.  This observation leads us to the following
definition of \emph{c-injective object}.
(The letter c here comes from \emph{Cartesian}.)

\begin{mydefinition}[c-injective object]
  \label{def:cinjectiveobject}
  Let $p\colon\EE\to\CC$ be a fibration.  An object $\bOmega\in\EE$ is
  a \emph{c-injective object} if the functor
  $\EE(-,\bOmega)\colon\EE^\op\to\Set$ sends every Cartesian arrow
  to a surjective map.

  Equivalently, $\bOmega\in\EE$ is a c-injective object if, for any
  Cartesian arrow $f\colon P\to Q$ in $\EE$ and any (not necessarily Cartesian) arrow
  $g\colon P\to \bOmega$, there is a (not necessarily Cartesian) arrow $h\colon Q\to \bOmega$
  satisfying $g=h\co f$.
\end{mydefinition}

Some basic objects can be shown to be c-injective objects.

\begin{myexample}[the two-point set]
  \label{exam:cinjective-equivalencerelation}
  In the fibration $\EqRel\to\Set$, $(2,=)$ is a c-injective object.
  Here, $2=\{\bot,\top\}$ is the two-point set and $=$ means the
  equality relation.  Indeed, for any Cartesian
  $f\colon (X,R_X)\to(Y,R_Y)$ and any $g\colon(X,R_X)\to(2,=)$, if we
  define $h\colon(Y,R_Y)\to(2,=)$ by
  \begin{equation*}
    h(y) =
    \begin{cases}
      g(x) & \text{if } (y, f(x))\in R_Y \\
      \top & \text{otherwise,}
    \end{cases}
  \end{equation*}
  then this turns out to be well-defined and satisfies $h\co f = g$.
\end{myexample}

\begin{myexample}[the two-point poset of truth values]
  \label{exam:cinjective-preorder-2}
  In the fibration $\Pre\to\Set$, $(2,\le)$ is a c-injective object.
  Here, $\le$ is the unique
  partial order satisfying $\bot\le\top$ and $\top\nleq\bot$.
  Indeed, for any Cartesian arrow $f\colon (X,R_X)\to(Y,R_Y)$ and any $g\colon(X,R_X)\to(2,\le)$,
  if we define $h\colon\colon(Y,R_Y)\to(2,\le)$ by
  \begin{equation*}
    h(y) =
    \begin{cases}
      \bot & \text{if } (y, f(x))\in R_Y \text{ for some } x \text{ such that } g(x)=\bot \\
      \top & \text{otherwise,}
    \end{cases}
  \end{equation*}
  then this turns out to be well-defined and satisfies $h\co f = g$.  
\end{myexample}

\begin{myexample}[{the unit interval as a pseudometric
    space~\cite[Theorem 5.8]{BaldanBKK-LMCS2018}}]
  \label{exam:cinjective-pseudometric}
  In the fibration $\PMetT\to\Set$, $[0,\top]$ is a c-injective
  object.  Indeed, for any arrow $g\colon (X,d_X)\to([0,\top],d_e)$
  and any Cartesian arrow $f\colon (X,d_X)\to(Y,d_Y)$, we can show
  that the map $h\colon Y\to[0,\top]$ defined by
  $h(y)=\inf_{x\in X}\left( g(x)+d_Y(f(x),y) \right)$ is nonexpansive
  from $(Y,d_Y)$ to $([0,\top],d_e)$.
\end{myexample}

The following non-example shows that c-injectivity crucially depends
on the fibration we consider.

\begin{myexample}[non-example]
  In contrast to \autoref{exam:cinjective-preorder-2}, in the
  fibration $\ERel\to\Set$, $(2,\le)$ is not c-injective, where
  $2=\{\bot,\top\}$ is the two-point set and $\le$ is the unique
  partial order satisfying $\bot\le\top$ and $\top\nleq\bot$.

  This can be seen as follows.  Let $X=\{a,b\}$, $Y=\{x,y,z\}$,
  $R_X=\emptyset$, and $R_Y=\{(x,z),(z,y)\}$.  Then $(X,R_X)$ and
  $(Y,R_Y)$ are objects of $\ERel$.  Consider the maps
  $f\colon (X,R_X)\to(Y,R_Y)$ and $g\colon(X,R_X)\to(2,\le)$ defined
  by $f(a)=x$, $f(b)=y$, $g(a)=\top$, and $g(b)=\bot$.  Note that $f$
  is Cartesian.  However, there is no $h\colon(Y,R_Y)\to(2,\le)$ such
  that $h\co f=g$: such $h$ would satisfy
  $\top = h(f(a))=h(x)\le h(z)\le h(y)= h(f(b))=\bot$, which
  contradict $\top\nleq\bot$.

  The same example can also be used to show that, in contrast to
  \autoref{exam:cinjective-equivalencerelation}, $(2,=)$ is not
  c-injective, where $=$ means the equality relation.
\end{myexample}

\subsection{Sufficient Condition for Fibered Codensity Lifting}

Now we are prepared to state the following main theorem of the current
paper.  The strategy of the proof is roughly as mentioned earlier.

\begin{mytheorem}[fiberedness from injective object]
  \label{thm:fiberednessfrominjective}
  In the setting of~\autoref{def:codensitylifting}, if $\bOmega$ is a
  c-injective object, then $\codlif{F}{\bOmega}{\tau}$ is fibered.
\end{mytheorem}
\begin{proof}
  Let $f\colon P\to Q$ be any Cartesian arrow.  By
  \autoref{prop:fiberednesspullback}, it suffices to show
  $\codlif{F}{\bOmega}{\tau}P = (F(pf))^* \left(
    \codlif{F}{\bOmega}{\tau}Q \right)$.  Here,
  $\codlif{F}{\bOmega}{\tau}P \lefib (F(pf))^* \left(
    \codlif{F}{\bOmega}{\tau}Q \right)$ has already been proven.
  Thus, our goal is the inequality
  $\codlif{F}{\bOmega}{\tau}P \gefib (F(pf))^* \left(
    \codlif{F}{\bOmega}{\tau}Q \right)$.

  Here, since $\bOmega$ is c-injective and $f$ is Cartesian, the
  following inclusion holds:
  \[ \left\{ g\in\EE(P,\bOmega) \right\} \subseteq \left\{ h\co f ~|~
      h\in\EE(Q,\bOmega)\right\}.
  \]
  By the definition of the meet, we have
  \[ \bigwedgefib_{g\in\EE(P,\bOmega)}(F(pg))^* \tau^* \bOmega \gefib
    \bigwedgefib_{h\in\EE(Q,\bOmega)} (F(p(h \co f)))^* \tau^*
    \bOmega.
  \]
  By the calculation in the proof of
  \autoref{prop:codensityliftingisfunctor}, this implies
  \[\codlif{F}{\bOmega}{\tau}P \gefib (F(pf))^* \left(
      \codlif{F}{\bOmega}{\tau}Q \right).\qedhere\]
\end{proof}

\begin{myremark}
  A refinement of \autoref{thm:fiberednessfrominjective} to an
  if-and-only-if result seems hard.  At least there is a simple
  counterexample to the most naive version of it: Consider a
  $\CLatw$-fibration $\Id\colon\CC\to\CC$, an endofunctor
  $\Id\colon\CC\to\CC$, an object $C\in\CC$, and an arrow
  $\tau\colon C\to C$.  The codensity lifting $\codlif{\Id}{C}{\tau}$
  is always equal to $\Id$, which is fibered.  However, since any
  arrow in $\CC$ is a Cartesian arrow w.r.t.~$\Id$, it is not hard to
  find an example of $C$ and $\CC$ such that $C$ is not c-injective
  w.r.t.~$\Id$.
\end{myremark}

\begin{myexample}[Kantorovich lifting]
  \label{exam:kantorovichlifting}
  Baldan et al.~\cite[Theorem 5.8]{BaldanBKK-LMCS2018} have shown that
  any Kantorovich lifting preserves isometries.  In terms of
  fibrations, this means that such functor is a fibered
  endofunctor on the fibration $\PMetT\to\Set$.
  
  Since Kantorovich lifting is a special case of codensity lifting
  where $\bOmega = ([0,\top],d_{\RR})$,
  \autoref{thm:fiberednessfrominjective} and
  \autoref{exam:cinjective-pseudometric} recover the same result.
  Actually, this has inspired \autoref{thm:fiberednessfrominjective}
  as a prototype.
\end{myexample}

The argument above also applies to situations with multiple parameters.

\begin{mydefinition}[codensity lifting with multiple parameters (as
  in~\cite{KomoridaKHKH-LICS2019})]
  \label{def:codensityliftingmulti}
  Let $\EE,\CC,p$, and $F$ be as in \autoref{def:codensitylifting}.
  Let $A$ be a set.  Assume that, for each $a\in A$, we are given
  $\bOmega_a \in\EE$ above $\Omega_a \in\CC$ and
  $\tau_a\colon F\Omega_a\to\Omega_a$.  Define a functor
  $\codlif{F}{\bOmega}{\tau}\colon\EE\to\EE$ by
  \[ \codlif{F}{\bOmega}{\tau}P=\bigwedgefib_{a\in
      A}\codlif{F}{\bOmega_a}{\tau_a}P
  \] for each $P\in\EE$.
\end{mydefinition}

\begin{mycorollary}
  \label{cor:fiberednessfrominjectivemulti}
  In the setting of \autoref{def:codensityliftingmulti}, if, for each
  $a\in A$, $\bOmega_a$ is a c-injective object, then
  $\codlif{F}{\bOmega}{\tau}$ is fibered.
\end{mycorollary}
\begin{proof}
  For any $P\in\EE$ above $X\in\CC$ and $f\colon Y\to X$ in $\CC$,
  using \autoref{thm:fiberednessfrominjective}, we
  can see
  \begin{align*}
    (Ff)^*\codlif{F}{\bOmega}{\tau}P &= (Ff)^*\bigwedgefib_{a\in A}\codlif{F}{\bOmega_a}{\tau_a}P& &= \bigwedgefib_{a\in A}(Ff)^*\codlif{F}{\bOmega_a}{\tau_a}P & & \\
                                     &= \bigwedgefib_{a\in A}\codlif{F}{\bOmega_a}{\tau_a}f^*P& &= \codlif{F}{\bOmega}{\tau}f^*P. &\qed&
  \end{align*}
\end{proof}

\begin{myexample}[Kantorovich lifting with multiple parameters]
  \label{exam:kantorovichliftingmulti}
  In~\cite{KoenigMikaMichalski-CONCUR2018}, K\"{o}nig and
  Mika-Michalski introduced a generalized version of Kantorovich
  lifting.
  
  Since it is a special case of \autoref{def:codensityliftingmulti}
  where $p$ is the fibration $\PMetT\to\Set$ and
  $\bOmega=([0,\top],d_{\RR})$,
  \autoref{cor:fiberednessfrominjectivemulti} and
  \autoref{exam:cinjective-pseudometric} imply that such lifting
  always preserves isometries.
\end{myexample}

\section{Results on C-injective Objects}
\label{sec:moreoncinjectives}

Here we seek properties of c-injective objects, mainly to obtain more
examples of them. We also see that, in a few fibrations, c-injective objects have been essentially identified by previous works.

\subsection{$\mclass$-injective Objects}
\label{sec:minjective}

To connect c-injectivity with existing works, we consider
a more general notion of $\mclass$-injective object.  The following
definition is found e.g.~in~\cite[Section
9.5]{KashiwaraSchapira-book2006}.
\begin{mydefinition}
  Let $\CC$ be a category and $\mclass$ be a class of arrows in
  $\CC$.  An object $X\in\CC$ is an \emph{$\mclass$-injective object}
  if the functor $\CC(-,X)\colon\CC^\op\to\Set$ sends every arrow
  in $\mclass$ to a surjective map.
\end{mydefinition}
The definition of c-injective objects is a special case of the definition above where $\mclass$ is the class of all Cartesian arrows.

The following is a folklore result.  The dual is found
e.g.~in~\cite[Proposition 10.2]{HiltonStammbach-book1997}.
\begin{myproposition}
  \label{prop:adjointinjective}
  Let $\CC,\DD$ be categories, $\mclass_\CC , \mclass_\DD$ be classes
  of arrows, and $L\dashv R\colon\CC\to\DD$ be a pair of adjoint
  functors.  Assume that $L$ sends any arrow in $\mclass_\DD$ to
  one in $\mclass_\CC$.  For any $\mclass_\CC$-injective $C\in\CC$,
  $RC\in\DD$ is $\mclass_\DD$-injective.
\end{myproposition}
\begin{proof}
  It suffices to show that $\DD(-,RC)\colon\DD^\op\to\Set$ sends each
  arrow in $\mclass_\DD$ to a surjective map.  By the assumption,
  the functor above factorizes to $L\colon\DD\to\CC$ and
  $\CC(-,C)\colon\DD^\op\to\Set$.  The former sends each arrow in
  $\mclass_\DD$ to one in $\mclass_\CC$ and the latter sends one in
  $\mclass_\CC$ to a surjective map.  Thus, the composition of these
  sends each arrow in $\mclass_\DD$ to a surjective map.\qed
\end{proof}

For epireflective subcategories, we have a sharper result:
\begin{myproposition}
  \label{prop:epireflectioninjective}
  In the setting of \autoref{prop:adjointinjective}, assume, in
  addition,
  \begin{itemize}
  \item $R$ is fully faithful,
  \item $R$ sends each arrow in $\mclass_\CC$ to one in
    $\mclass_\DD$, and
  \item each component of the unit $\eta\colon\Id\to RL$ is an
    epimorphism in $\mclass_\DD$.
  \end{itemize}
  Then, $D\in\DD$ is $\mclass_\DD$-injective if and only if it is
  isomorphic to $RC$ for some $\mclass_\CC$-injective $C\in\CC$.
\end{myproposition}
\begin{proof}
  The ``if'' part is \autoref{prop:adjointinjective}.  We show the
  ``only if'' part.
  
  Let $D\in\DD$ be any $\mclass_\DD$-injective object.  Since
  $\eta_D\colon D\to RLD$ is in $\mclass_\DD$, we can use the
  $\mclass_\DD$-injectiveness of $D$ to obtain $f\colon RLD\to D$ such
  that $f\co\eta_D = \id_D$.  Here, $\eta_D \co f \co \eta_D = \eta_D$
  and, by epi-ness of $\eta_D$, $\eta_D \co f = \id_{RLD}$.  Thus,
  $\eta_D$ is an isomorphism.

  Now we show that $LD$ is $\mclass_\CC$-injective.  Let
  $f\colon C\to LD$ and $g\colon C\to C'$ be any arrow in $\CC$
  and assume that $g$ is in $\mclass_\CC$.  Send these by $R$ to $\DD$
  and consider $Rf$ and $Rg$.  By the assumption, $Rg$ is in
  $\mclass_\DD$.  Since $RLD$ is isomorphic to $D$, it is also
  $\mclass_\DD$-injective.  Using these, we can obtain
  $h'\colon RC'\to RLD$ such that $h'\co Rg=Rf$.  Since $R$ is full,
  there is $h\colon C'\to LD$ such that $Rh=h'$.  The faithfulness of $R$
  implies $h\co g=f$.  Thus $LD$ is $\mclass_\CC$-injective.\qed
\end{proof}

Using this result, we can identify c-injective objects in a few
situations.
\begin{myexample}[continuous lattices in
  $\Top\to\Set$~\cite{Scott-1972-contlattice}]
  \label{exam:cinjective-topology}
  In the setting of \autoref{prop:epireflectioninjective}, consider
  the case where $\DD=\Top$, $\CC=\Top_0$.  Here $\Top_0$ is the full
  subcategory of $\Top$ of $T_0$ spaces.
  Let $R$ be the inclusion. It has a left adjoint $L$, taking each space to its Kolmogorov quotient.
  Let $\mclass_\CC$ be the
  class of topological embeddings (i.e. homeomorphisms to their images) and $\mclass_\DD$ be the class of Cartesian
  arrows (w.r.t. the fibration $\Top\to\Set$).  Then the
  assumptions in \autoref{prop:epireflectioninjective} are satisfied
  and we can conclude that c-injective objects in $\Top$ are precisely
  injective objects in $\Top_0$ w.r.t. embeddings.

  The latter has been identified by
  Scott~\cite{Scott-1972-contlattice}.  According to his result, such
  objects are precisely \emph{continuous lattices} with the Scott
  topology.  Thus, we can see that c-injective objects in $\Top$ are
  precisely such spaces.
\end{myexample}

\begin{myexample}[complete lattices in
  $\Pre\to\Set$~\cite{BanaschewskiBruns-1967}]
  \label{exam:cinjective-preorder}
  In the setting of \autoref{prop:epireflectioninjective}, consider
  the case where $\DD=\Pre$, $\CC=\Pos$.  Here $\Pos$ is the full
  subcategory of $\Pre$ of posets.
  Let $R$ be the inclusion. It has a left adjoint $L$, taking each preordered set to its poset reflection.
  Let $\mclass_\CC$ be the class of
  embeddings and $\mclass_\DD$ be the class of Cartesian arrows
  (w.r.t. the fibration $\Pre\to\Set$).  Then the assumptions in
  \autoref{prop:epireflectioninjective} are satisfied and we can
  conclude that c-injective objects in $\Pre$ are precisely injective
  objects in $\Pos$ w.r.t. embeddings.

  The latter has been identified by Banaschewski and
  Bruns~\cite{BanaschewskiBruns-1967} .  According to their result,
  such objects are precisely complete lattices.  Thus, we can see that
  c-injective objects in $\Pre$ are precisely complete lattices.
\end{myexample}

\subsection{Results Specific to C-injective Objects}

To develop the theory of c-injective objects further, we establish some preservation results for c-injectivity.
Based on the two propositions of the last section, we show two
propositions specific to fibrations and c-injective objects.

From \autoref{prop:adjointinjective}, we can derive the following:
\begin{myproposition}
  \label{prop:fibadjointcinjective}
  Let $p\colon\EE\to\CC,q\colon\FF\to\DD$ be $\CLatw$-fibrations and
  $L\dashv R\colon\EE\to\FF$ be a pair of adjoint functors.  If $L$ is
  fibered (from $q$ to $p$), then $RE\in\FF$ is c-injective (in $q$)
  for each c-injective $E\in\EE$.
\end{myproposition}
\begin{proof}
  Let $\mclass_\EE$ be the class of all arrows Cartesian w.r.t.~$p$ and
  $\mclass_\FF$ be the class of all arrows Cartesian w.r.t.~$q$.
  Then, use \autoref{prop:adjointinjective} to the pair $L\dashv R$ of
  adjoint functors.\qed
\end{proof}

From \autoref{prop:epireflectioninjective}, we can derive the following:
\begin{myproposition}
  \label{prop:fibepireflectioncinjective}
  In the setting of \autoref{prop:fibadjointcinjective}, assume in addition
  that both $L$ and $R$ are fibered and that $\eta\colon\Id\to RL$ is
  componentwise epi.  Then, $F\in\FF$ is c-injective if and only if it
  is isomorphic to $RE$ for some c-injective $E\in\EE$.
\end{myproposition}
\begin{proof}
  Use \autoref{prop:epireflectioninjective} in the same setting as the
  proof of \autoref{prop:fibadjointcinjective}.\qed
\end{proof}

\section{Examples}
\label{sec:examples}

We list several examples of \autoref{thm:fiberednessfrominjective}.
Indeed, most of the examples listed in~\cite[Table
VI]{KomoridaKHKH-LICS2019} turn out to be fibered by
\autoref{thm:fiberednessfrominjective}.  Since the conditions in
\autoref{thm:fiberednessfrominjective} only refer to
$p\colon\EE\to\CC$ and $\bOmega$, we sort the examples by these data.

We here recall some basic functors considered:
\begin{mydefinition}
  Let $\pow\colon\Set\to\Set$ be the covariant powerset functor and
  $\Distle\colon\Set\to\Set$ be the subdistribution functor.  Here, a
  subdistribution $p\in\Distle X$ is a measure on the $\sigma$-algebra
  of all subsets of $X$ with total mass $\le 1$.  We abbreviate
  $p(\{x\})$ to $p(x)$.
\end{mydefinition}

\subsection{Kantorovich Lifting}

In \autoref{exam:cinjective-pseudometric} we have seen that, in the
fibration $\PMetT\to\Set$, the object $([0,\top],d_{\RR})$ is
c-injective.  We gather examples of this case here.  As mentioned
in \autoref{exam:kantorovichlifting} and
\autoref{exam:kantorovichliftingmulti}, this class of examples has
been already studied and shown to be fibered
in~\cite{BaldanBKK-LMCS2018,KoenigMikaMichalski-CONCUR2018}.

\begin{myexample}[Hausdorff pseudometric]
  Let $\inf\colon\pow[0,\top]\to[0,\top]$ be the map taking any set to
  its infimum.  Then, the codensity lifting
  $\codlif{\pow}{([0,\top],d_\RR)}{\inf}\colon\PMetT\to\PMetT$ turns
  out to induce the \emph{Hausdorff distance}: for any
  $(X,d_X)\in\PMetT$, if we let
  $(\pow X,d_{\pow X})=\codlif{\pow}{([0,\top],d_\RR)}{\inf}(X,d_X)$,
  then
  \[ d_{\pow X}(S,T) = \max \left( \sup_{x\in S}\inf_{y\in T}d_X(x,y)
      , \sup_{y\in T}\inf_{x\in S}d_X(x,y)\right)
  \] holds for any $S,T\in\pow X$.  By
  \autoref{thm:fiberednessfrominjective}, this functor is fibered.
\end{myexample}

\begin{myexample}[Kantorovich pseudometric]
  Let $e\colon\Distle[0,\top]\to[0,\top]$ be the map taking any
  distribution to its expected value. Then, the codensity lifting
  \[
    \codlif{\Distle}{([0,\top],d_\RR)}{e}\colon\PMetT\to\PMetT
  \] turns out to induce the \emph{Kantorovich distance}: for any
  $(X,d_X)\in\PMetT$, if we let
  $(\Distle X,d_{\Distle
    X})=\codlif{\Distle}{([0,\top],d_\RR)}{e}(X,d_X)$, then
  \[ d_{\Distle X}(p,q) = \sup_{f\colon (X,d_X)\to
      ([0,\top],d_\RR)\text{ nonexpansive}} \left| \sum_{x\in
        X}f(x)p(x)-\sum_{x\in X}f(x)q(x)\right|
  \] holds for any $p,q\in\Distle X$.  By
  \autoref{thm:fiberednessfrominjective}, this functor is fibered.
\end{myexample}

\subsection{Lower, Upper, and Convex Preorders}

In \autoref{exam:cinjective-preorder}, we have identified complete lattices as c-injective
objects in the fibration $\Pre\to\Set$.  In particular, the two-point
set $(2,\le)$ is a c-injective object
(\autoref{exam:cinjective-preorder-2}).

Katsumata and Sato~\cite[Section 3.1]{KatsumataSato-CALCO2015} used codensity
lifting to recover the \emph{lower}, \emph{upper}, and \emph{convex
  preorders} on powersets.  Here we see that our result applies to
them: all of the following liftings are fibered.

\begin{myexample}[lower preorder]
  \label{exam:lowerpreorder}
  Define $\Diamond\colon\pow 2\to 2$ so that $\Diamond S=\top$ if and
  only if $\top\in S$.  Then, the codensity lifting
  $\codlif{\pow}{(2,\le)}{\Diamond}\colon\Pre\to\Pre$ turns out to
  induce the \emph{lower preorder}: if we let
  $(\pow X,\le_{\pow
    X}^{\Diamond})=\codlif{\pow}{(2,\le)}{\Diamond}(X,\le_X)$, then,
  for any $S,T\in\pow X$,
  \[ S\le_{\pow X}^{\Diamond}T \Leftrightarrow \forall x\in S, \exists
    y\in T,x\le_X y.
  \]
\end{myexample}

\begin{myexample}[upper preorder]
  \label{exam:upperpreorder}
  Define $\square\colon\pow 2\to 2$ so that $\square S=\top$ if and
  only if $\bot\notin S$.  Then, the codensity lifting
  $\codlif{\pow}{(2,\le)}{\square}\colon\Pre\to\Pre$ turns out to
  induce the \emph{upper preorder}: if we let
  $(\pow X,\le_{\pow
    X}^{\square})=\codlif{\pow}{(2,\le)}{\square}(X,\le_X)$, then,
  for any $S,T\in\pow X$,
  \[ S\le_{\pow X}^{\square}T \Leftrightarrow \forall y\in T, \exists
    x\in S,x\le_X y.
  \]
\end{myexample}

\begin{myexample}[convex preorder]
  \label{exam:convexpreorder}
  Denote the family of the two lifting parameters above by
  $((2,\le),\{\Diamond,\square\})$.  Then, the codensity lifting (with
  multiple parameters, \autoref{def:codensityliftingmulti})
  $\codlif{\pow}{(2,\le)}{\{\Diamond,\square\}}\colon\Pre\to\Pre$ is
  simply the meet of $\codlif{\pow}{(2,\le)}{\Diamond}$ and
  $\codlif{\pow}{(2,\le)}{\square}$. This is what is called the
  \emph{convex preorder}.
\end{myexample}

\begin{myremark}
  The original formulation~\cite[Section 3.1]{KatsumataSato-CALCO2015}
  is based on codensity lifting of monads, so apparently different to
  ours.  In our terms, they used the multiplication
  $\mu_1\colon\pow\pow 1 \to \pow 1$ and two different preorders on
  $\pow 1$.  Using two different bijections between $\pow 1$ and $2$,
  it can be shown that their formulation is actually equivalent to
  ours.
\end{myremark}

\subsection{Equivalence relations}
In \autoref{exam:cinjective-equivalencerelation} we have seen that, in
the fibration $\EqRel\to\Set$, the object $(2,=)$ is c-injective.  We
gather examples of this case here. All of the following liftings
are fibered. Details on the following examples can be found
in~\cite{KomoridaKHKH-LICS2019}.

\begin{myexample}[lifting for bisimilarity on Kripke frames]
  Consider the codensity lifting
  $\codlif{\pow}{(2,=)}{\Diamond}\colon\EqRel\to\EqRel$, where
  $\Diamond$ is as defined in \autoref{exam:lowerpreorder}.  This
  turns out to satisfy the following: if we let
  $(\pow X,\sim_{\pow X})=\codlif{\pow}{(2,=)}{\Diamond}(X,\sim_X)$,
  then
  \[ S\sim_{\pow X}T \Leftrightarrow \left( \forall x\in S, \exists
      y\in T,x\sim_X y \right) \wedge \left( \forall y\in T, \exists
      x\in S,x\sim_X y \right)
  \] holds for any $S,T\in\pow X$.  This can be used to define (the
  conventional notion of) bisimilarity on Kripke frames
  ($\pow$-coalgebras).
\end{myexample}

\begin{myexample}[lifting for bisimilarity on Markov chains]
  For each $r\in [0,1]$, define a map $\Thr_r\colon\Distle 2\to 2$ so
  that $\Thr_r(p)=\top$ if and only if $p(\top)\ge r$. These define a
  $[0,1]$-indexed family of lifting parameters
  $((2,=),\Thr_r)_{r\in[0,1]}$. The codensity lifting
  $\codlif{\Distle}{(2,=)}{\Thr}$ defined by this family can be used
  to define probabilistic bisimilarity on Markov chains
  ($\Distle$-coalgebras).
\end{myexample}

\subsection{Topologies}

In \autoref{exam:cinjective-topology}, we have identified c-injective objects in the fibration $\Top\to\Set$.
In particular, the \emph{Sierpinski space}, defined as follows, is a c-injective object:
\begin{mydefinition}[Sierpinski space]
  The \emph{Sierpinski space} is a topological space
  $(2,\OpenSet_\Sierp)$ where $2=\{\bot,\top\}$ and the family
  $\OpenSet_\Sierp$ of open sets is $\{\emptyset,\{\top\},2\}$.
  We denote this space by $\Sierp$.
\end{mydefinition}

The following liftings of $\pow$ have appeared in~\cite[Section
3.2]{KatsumataSato-CALCO2015}. All of them are fibered: in other
words, they send embeddings to embeddings.
\begin{myexample}[lower Vietoris lifting]
  Consider the codensity lifting
  $\codlif{\pow}{\Sierp}{\Diamond}\colon\Top\to\Top$, where $\Diamond$
  is as defined in \autoref{exam:lowerpreorder}.  For each
  $(X,\OpenSet_X)\in\Top$, if we let
  $(\pow X,\OpenSet_{\pow
    X}^{\Diamond})=\codlif{\pow}{\Sierp}{\Diamond}(X,\OpenSet_X)$,
  then the topology $\OpenSet_{\pow X}^{\Diamond}$ is the coarsest one
  such that, for each $U\in\OpenSet_X$, the set
  $\{V\subseteq X~|~V\cap U\neq\emptyset\}$ is open.  This is called
  \emph{lower Vietoris lifting} in~\cite{KatsumataSato-CALCO2015}.
\end{myexample}

\begin{myexample}[upper Vietoris lifting]
  Consider the codensity lifting
  $\codlif{\pow}{\Sierp}{\square}\colon\Top\to\Top$, where $\square$
  is as defined in \autoref{exam:upperpreorder}.  For each
  $(X,\OpenSet_X)\in\Top$, if we let
  $(\pow X,\OpenSet_{\pow
    X}^{\square})=\codlif{\pow}{\Sierp}{\square}(X,\OpenSet_X)$, then
  the topology $\OpenSet_{\pow X}^{\square}$ is the coarsest one such
  that, for each $U\in\OpenSet_X$, the set
  $\{V\subseteq X~|~V\subseteq U\}$ is open.  This is called
  \emph{upper Vietoris lifting} in~\cite{KatsumataSato-CALCO2015}.
\end{myexample}

\begin{myexample}[Vietoris lifting]
  Define the codensity lifting
  $\codlif{\pow}{\Sierp}{\{\Diamond,\square\}}\colon\Top\to\Top$ like
  one in \autoref{exam:convexpreorder}.  We call this \emph{Vietoris
    lifting}.
  
  This turns out to be connected to \emph{Vietoris
    topology}~\cite{KupkeKV-ENTCS2003} as follows.  For each
  $(X,\OpenSet_X)\in\Top$, let
  $(\pow X,\OpenSet_{\pow
    X}^{\Diamond,\square})=\codlif{\pow}{\Sierp}{\{\Diamond,\square\}}(X,\OpenSet_X)$.
  The set $K(X,\OpenSet_X)$ of closed subsets of $(X,\OpenSet_X)$ is a
  subset of $\pow X$.  Here, the topology on $K(X,\OpenSet_X)$ induced
  from $\OpenSet_{\pow X}^{\Diamond,\square}$ is the same as the
  Vietoris topology.

  This coincidence and the fiberedness of
  $\codlif{\pow}{\Sierp}{\{\Diamond,\square\}}$ implies that the
  \emph{Vietoris functor}
  $\mathbb{V}\colon\mathbf{Stone}\to\mathbf{Stone}$, defined
  in~\cite{KupkeKV-ENTCS2003}, sends embeddings to embeddings.
\end{myexample}

In~\cite{KomoridaKHKH-LICS2019}, we considered another lifting:
\begin{myexample}[lifting for bisimulation topology]
  \label{exam:bisimulationtopology}
  Fix any set $\Sigma$.  Let $A_\Sigma\colon\Set\to\Set$ be the
  functor defined by $A_\Sigma X = 2\times X^\Sigma$.  Define
  $\Acc\colon A_\Sigma 2\to 2$ by $\Acc(t,\rho)=t$.  For each
  $a\in\Sigma$, define $\Next{a}\colon A_\Sigma 2\to 2$ by
  $\Next{a}(t,\rho)=\rho(a)$.  Here, $(\Sierp,\Acc)$ and
  $(\Sierp,\Next{a})$ for each $a\in\Sigma$ consist of a family of
  lifting parameters.  The codensity lifting (with multiple parameters,
  \autoref{def:codensityliftingmulti})
  $\codlif{A_\Sigma}{\Sierp}{\{\Acc\}\cup\{\Next{a}|a\in\Sigma\}}\colon\Top\to\Top$
  was used to define \emph{bisimulation topology} for deterministic
  automata ($A_\Sigma$-coalgebras). This is fibered.
  This fact is used in \autoref{exam:bisimtopfromfinalcoalg}, where we will look at bisimulation topology again.
\end{myexample}

\section{Application to Codensity Bisimilarity}
\label{sec:applications}

Now we present an application of our main result.  Based on codensity
lifting, we defined \emph{codensity bisimilarity}
in~\cite{KomoridaKHKH-LICS2019}.  It subsumes bisimilarity, simulation
preorder, and behavioral metric as special cases.  Here we see that,
in the cases to which our fiberedness result applies, codensity bisimilarity
interacts well with coalgebra morphisms.  In particular, the codensity
bisimilarity on any coalgebra is determined by that on the final
coalgebra, if it exists.

Recall the definition of coalgebra:
\begin{mydefinition}[coalgebra of an endofunctor]
  Let $F\colon\CC\to\CC$ be an endofunctor on a category $\CC$.  An
  \emph{$F$-coalgebra} is a pair of an object $X\in\CC$ and an arrow
  $c\colon X\to FX$.

  Let $c\colon X\to FX$ and $d\colon Y\to FY$ be $F$-coalgebras.  A
  \emph{morphism of coalgebras} from $(X,c)$ to $(Y,d)$ is an arrow
  $f\colon X\to Y$ in $\CC$ such that $d\co f= Ff\co c$ holds.
\end{mydefinition}

As sketched in Section~\ref{sec:introduction}, functor lifting can be used to define a ``bisimilarity-like notion''.
If we use codensity lifting in this construction, we obtain the following definition:
\begin{mydefinition}[{codensity bisimilarity~\cite[Definitions III.6
    and III.8]{KomoridaKHKH-LICS2019}}]
  Assume the setting of \autoref{def:codensityliftingmulti}.  Let
  $c\colon X\to FX$ be any $F$-coalgebra.  Define
  $\codPT{\bOmega}{\tau}{c}\colon\EE_X\to\EE_X$ by
  $\codPT{\bOmega}{\tau}{c}P =
  c^*\left(\codlif{F}{\bOmega}{\tau}P\right)$.

  The \emph{($(\bOmega,\tau)$-)codensity bisimilarity} is the greatest
  fixed point (w.r.t.$\lefib$) of $\codPT{\bOmega}{\tau}{c}$.  We
  denote this by $\nu\codPT{\bOmega}{\tau}{c}$.
\end{mydefinition}

Note that the greatest fixed point of $\codPT{\bOmega}{\tau}{c}$
always exists.  This can be seen, for example, by the Tarski fixed
point theorem.  Another option is to use the constructive fixed point
theorem by Cousot and Cousot~\cite{CousotCousot-1979-fixedpoint}.  We
use their characterization of the greatest fixed point to prove the
following proposition:

\begin{myproposition}[stability of codensity bisimilarity]
  \label{prop:coalgebramorphismiscartesian}
  Assume the setting of \autoref{def:codensityliftingmulti} (codensity lifting with multiple parameters).
  Assume also that each $\bOmega_a$ is a c-injective object.
  Then, codensity bisimilarity is stable under coalgebra morphisms:
  for any morphism of coalgebras $f$ from $(X,c)$ to $(Y,d)$,
  we have $\nu\codPT{\bOmega}{\tau}{c} = f^*\left(\nu\codPT{\bOmega}{\tau}{d}\right)$.
\end{myproposition}
\begin{proof}
  Define a transfinite sequence
  $\left(\nu_\alpha\codPT{\bOmega}{\tau}{c}\right)_{\alpha\text{ is an ordinal}}$ of elements
  of $\EE_X$ by the
  following:\[ \nu_\alpha\codPT{\bOmega}{\tau}{c} =
    \bigwedgefib_{\beta<\alpha}\codPT{\bOmega}{\tau}{c}\left(\nu_\beta\codPT{\bOmega}{\tau}{c}\right).
  \]
  Define another transfinite sequence
  $\left(\nu_\alpha\codPT{\bOmega}{\tau}{d}\right)_{\alpha\text{ is an
      ordinal}}$ by a similar manner.  By the result
  in~\cite{CousotCousot-1979-fixedpoint}, there is an ordinal $\gamma$
  such that
  $\nu_\gamma\codPT{\bOmega}{\tau}{c}=\nu\codPT{\bOmega}{\tau}{c}$ and
  $\nu_\gamma\codPT{\bOmega}{\tau}{d}=\nu\codPT{\bOmega}{\tau}{d}$.
  \footnote{This formulation differs slightly from the conventional one where successor and limit ordinals are distinguished, but the result also holds under this definition.}
  Thus, it suffices to show the following claim:
  \begin{myclaim}
    For any ordinal $\alpha$, we have $\nu_\alpha\codPT{\bOmega}{\tau}{c} =
      f^*\left(\nu_\alpha\codPT{\bOmega}{\tau}{d}\right)$.
  \end{myclaim}
  We show this by transfinite induction on $\alpha$.
  Assume the claim holds for all $\beta<\alpha$.

  Using the assumption that $f$ is a morphism of coalgebras, the
  fiberedness of $\codlif{F}{\bOmega}{\tau}$
  (\autoref{cor:fiberednessfrominjectivemulti}), and the functoriality
  of pullback (\autoref{prop:pullbackisfunctorial}), we have
  $f^*\co\codPT{\bOmega}{\tau}{d}=\codPT{\bOmega}{\tau}{c}\co f^*$.
  It implies the claim for $\alpha$
  \begin{align*}
    f^*\left(\nu_\alpha\codPT{\bOmega}{\tau}{d}\right) &= f^*\left(\bigwedgefib_{\beta<\alpha}\codPT{\bOmega}{\tau}{d}\left(\nu_\beta\codPT{\bOmega}{\tau}{d}\right)\right) 
                                                       = \bigwedgefib_{\beta<\alpha}f^*\left(\codPT{\bOmega}{\tau}{d}\left(\nu_\beta\codPT{\bOmega}{\tau}{d}\right)\right) & & \\
                                                       &= \bigwedgefib_{\beta<\alpha}\codPT{\bOmega}{\tau}{c}\left(f^*\nu_\beta\codPT{\bOmega}{\tau}{d}\right) 
                                                       = \bigwedgefib_{\beta<\alpha}\codPT{\bOmega}{\tau}{c}\left(\nu_\beta\codPT{\bOmega}{\tau}{c}\right) & & \\
                                                       &= \nu_\alpha\codPT{\bOmega}{\tau}{c}. & & &\qed& \\
  \end{align*}
\end{proof}

In particular, the codensity bisimilarity is determined by that on the final coalgebra:
\begin{mycorollary}
  \label{cor:codenbisimfromfinalcoalg}
  Assume the setting of \autoref{prop:coalgebramorphismiscartesian}.
  Assume also that there exists a final $F$-coalgebra
  $z\colon Z\to FZ$.  Then, for any $F$-coalgebra $c\colon X\to FX$,
  the unique coalgebra morphism $!_X\colon X\to Z$ satisfies
  $\nu\codPT{\bOmega}{\tau}{c} =
    (!_X)^*\left(\nu\codPT{\bOmega}{\tau}{z}\right)$.
\end{mycorollary}

\begin{myexample}[bisimulation topology for deterministic automata]
  \label{exam:bisimtopfromfinalcoalg}
  Recall \autoref{exam:bisimulationtopology}.  For any
  $A_\Sigma$-coalgebra $c\colon X\to A_\Sigma X$, we defined the
  codensity bisimilarity on $X$ by
  $\nu\codPT{\Sierp}{\{\Acc\}\cup\{\Next{a}|a\in\Sigma\}}{c}\in\Top_X$~\cite{KomoridaKHKH-LICS2019}.

  The functor $A_\Sigma$ has a final coalgebra: the set $2^{\Sigma^*}$
  of all languages on the alphabet $\Sigma$ can be given an
  $A_\Sigma$-coalgebra structure and it is final.  For an
  $A_\Sigma$-coalgebra $c\colon X\to A_\Sigma X$, the unique coalgebra
  morphism $l\colon X\to 2^{\Sigma^*}$ assigns to each state the
  recognized language when started from it.

  \autoref{cor:codenbisimfromfinalcoalg} implies that this map $l$
  determines the bisimulation topology on $X$.  We believe that this
  fact is new, and it supports our use of the term \emph{language
    topology} in~\cite[\S VIII-C]{KomoridaKHKH-LICS2019}.
\end{myexample}

\section{Conclusions}
\label{sec:conclusions}

Inspired by the proof of fiberedness of Kantorovich
lifting~\cite{BaldanBKK-LMCS2018}, we showed a sufficient condition
for codensity lifting to be fibered.  We listed a number of examples
that satisfy the sufficient condition.  In addition, we apply the
fiberedness to show a result on codensity bisimilarity.

One possible direction of research is to investigate the notion of
c-injectiveness in more depth.  The existing work on injective
objects in homological algebra and topos theory can be a clue for that.  In
particular, we have not studied which category has \emph{enough
  c-injectives}.  This may be connected with some deep fibrational
property.

Another possible direction is to generalize the main result.
In~\cite{KatsumataSato-CALCO2015}, codensity lifting of a monad was
introduced for a general fibration in terms of right Kan extension.
This definition can readily be adapted to endofunctors, but in the
current paper, we considered only $\CLatw$-fibrations.  Extending the
main result to this general situation, in particular, to non-poset
fibrations, may broaden the scope of application.
It can also be fruitful to extend the definition of codensity lifting itself: for example, in \autoref{def:codensitylifting}, we could substitute $\tau^*\bOmega$ with other objects above $F\Omega$.
\footnote{This has been pointed out by an anonymous reviewer.}
Seeking consequences and examples of this version of the definition is future work.
Another related research direction is to obtain a similar sufficient condition for fiberedness of \emph{categorical $\top\top$-lifting}~\cite{Katsumata-CSL2005}.

Last but not least, we have to seek other applications.  As mentioned
in Section~\ref{sec:introduction}, functor lifting is used in many
situations.  Using codensity lifting there and seeing what can be
implied by the current result seems to be a promising research
direction.  In particular, codensity lifting seems to be intimately
connected to \emph{coalgebraic modal logic}, where
$\tau\colon F\Omega\to\Omega$ is regarded as a \emph{modality}.
Recently, Kupke and Rot~\cite{KupkeRot-toappearCSL2020} have
identified a sufficient condition for a logic to expressive w.r.t. a
coinductive predicate (like bisimilarity, behavioral metric
etc.). They used fiberedness of lifting in a crucial way (they use the
term \emph{fibration map}), which suggests that the current work can
play a pivotal role in investigating modal logics.

\subsubsection{Acknowledgments}
The author is grateful to Ichiro Hasuo and Shin-ya Katsumata for
fruitful discussions on technical and structural points.  The author
is also indebted to anonymous reviewers for clarifying things and
pointing out possible future directions, including a topos-theoretic
viewpoint and an alternative definition of codensity lifting.

\bibliographystyle{splncs04} \bibliography{mybibliography}

\end{document}